\documentclass[twocolumn,preprintnumbers,amsmath,amssymb,dvipdfmx]{revtex4-1}

\usepackage{graphicx}
\usepackage{dcolumn}
\usepackage{bm}
\usepackage{color}
\usepackage[normalem]{ulem}

\begin{document}

\title{
STM as a single Majorana detector of Kitaev's chiral spin liquid
}

\author{Masafumi Udagawa$^{1,2}$, Shintaro Takayoshi$^{2,3}$ and Takashi Oka$^{2,4,5}$}%
\affiliation{%
$^1$Department of Physics, Gakushuin University, Mejiro, Toshima-ku, Tokyo 171-8588, Japan\\
$^2$Max-Planck-Institut f\"{u}r Physik komplexer Systeme, 01187 Dresden, Germany\\
$^3$Department of Physics, Konan University, Kobe, 658-8501, Japan\\
$^4$Institute for Solid State Physics, University of Tokyo, Kashiwa 277-8581, Japan\\
$^5$Trans-scale Quantum Science Institute, University of Tokyo, Bunkyo-ku, Tokyo 113-0033, Japan
}%

\date{\today}

\begin{abstract}
In this letter, we propose a local detection scheme for the Majorana zero mode (MZM) carried by a vison in Kitaev's chiral spin liquid (CSL) using scanning tunneling microscopy (STM). 
The STM introduces a single Majorana into the system through hole/charge injection and the Majorana interacts with the MZM to form a stable composite object.
We derive the exact analytical expression of single-hole Green's function in the Mott insulating limit of Kitaev's model, and 
show that the differential conductance has split peaks, as a consequence of resonant tunneling through the vison-hole composite.
The peak splitting scales with the binding energy of vison-hole composite, which is comparable to the Majorana gap in CSL, well within the reach of experimental observation. 
\end{abstract}

\maketitle
{\it Introduction}: In recent years, a number of candidate materials have been proposed for quantum spin liquids (QSL's), and intensive studies are going on for its realization~\cite{Balents:2010aa}. 
An important class of QSL phases is characterized by their topological nature of ground states and excitations~\cite{wen2004quantum,pachos2012introduction}.
On one hand, the underlying topological structure means the robustness of the phase, as local perturbations cannot change the global topology immediately.
However, on the other hand, due to its inherent non-locality, the topological structure sometimes disables direct access by local experimental probes and makes it a challenging task to identify the QSL phase experimentally.

Among many QSL candidates proposed so far, Kitaev's chiral spin liquid (CSL) state deserves special attention~\cite{kitaev2006anyons}.
The appearance of the CSL phase is theoretically predicted for Kitaev's honeycomb model in a magnetic field.
This phase hosts a finite Chern number, which results in half-integer quantization of thermal Hall conductivity, as recently claimed in the field-induced nonmagnetic state~\cite{Banerjee:2018aa,PhysRevB.91.180401} of $\alpha$-RuCl$_3$~\cite{Kasahara:2018aa,yamashita2020sample,yokoi2020half}. Moreover, due to the finite Chern number of the phase, the $Z_2$ vortex excitation called a vison, is turned into a non-Abelian Ising anyon accompanied with a Majorana zero mode (MZM).

Detection and control of MZM are of considerable interest in terms of both material physics and quantum information technology~\cite{KITAEV20032,nayak2008non}, and great efforts have been made for the identification of MZM in quantum wires~\cite{kitaev2001unpaired,Mourik1003,Nadj-Perge602,Albrecht:2016aa}, surface states of $^{3}$He and topological insulators~\cite{doi:10.1143/JPSJ.80.013602,Ikegami:2019aa,PhysRevLett.118.145301,PhysRevLett.100.096407}, topological superconductors~\cite{PhysRevLett.86.268,PhysRevB.79.094504,PhysRevB.82.134521,PhysRevLett.105.217001,PhysRevLett.104.040502,PhysRevLett.105.077001,PhysRevB.82.094504,PhysRevB.92.134519,PhysRevLett.111.136401,PhysRevB.94.060507}, non-Abelian Fractional Hall states~\cite{MOORE1991362,PhysRevLett.96.016803,Willett8853}, and so on.
Despite these keen interests, its detection involves fundamental difficulty; it is difficult to pick up a single MZM, as no local observables are coupled with a spatially isolated Majorana. 
However, in this work, we argue that hole/charge injection sheds a new light on this problem.

The interaction between an injected hole/charge and vison remains largely unexplored in the CSL. Recently, mobile carrier doping has been attempted for $\alpha$-RuCl$_3$~\cite{PhysRevMaterials.1.052001,PhysRevB.99.245141,Wang:2000aa,PhysRevB.100.165426,Mashhadi:2019aa}. Theoretically, the possibility of binding a vison to a static hole (vacancy) has been discussed in the anisotropic A phase and gapless B phase of Kitaev's spin liquid and related models~\cite{willans2010disorder,PhysRevB.84.115146,PhysRevB.90.035145,PhysRevB.94.235105,PhysRevB.98.220404,sanyal2020emergent}.

In this work,
we propose that a hole/charge injection serves as a sensitive method to detect the presence of a single MZM accompanying a vison; the injected hole/charge effectively introduces a single Majorana into the system, which makes a bonding state with the precedent MZM accompanied with a vison. This bonding state is much more robust, compared with the vison-vacancy composites
in A and B phases; the binding energy is comparable to the scale of the Majorana gap of CSL, i.e., the composite is well stabilized, as soon as the CSL phase is realized with a well-defined excitation gap.
To address the experimental consequence of this composite formation, we derived an exact analytical expression of hole Green's function at the Mott insulating limit of Kitaev's model and identified a low-energy peak in the local density of state, which indicates the presence of vison-hole composite. By importing this local density of states into the general formula for STM, we found the enhancement of differential conductance due to the resonant tunneling through the vison-hole composite, which provides a clear diagnose of the presence of MZM in the CSL phase.

\begin{figure}[h]
\begin{center}
\includegraphics[width=0.99\linewidth]{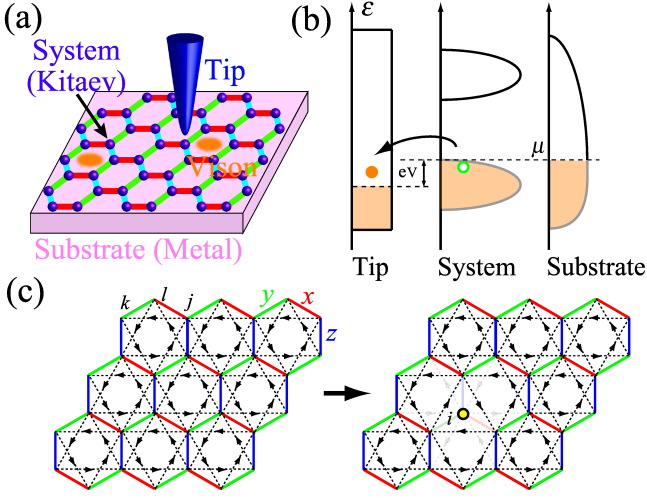}
\vspace{-10pt}
\caption{(Color online) (a)\ Theoretical setup of STM device. Electric current conducts from an STM tip to a metallic substrate through the Kitaev's system. (b)\ Schematic plot of density of states and electron occupation. The Kitaev's system is in equilibrium with the substrate with a common chemical potential, $\mu$, which is fixed to the top of lower Hubbard band of Kitaev Mott insulator. 
Electrons tunnel from the system to the tip, driven by the voltage bias, $V$. (c)\ Lattice convention of Kitaev's Hamiltonian. Red, green and blue bonds represent the Ising couplings of $x$, $y$ and $z$ components. The dashed lines express the hopping direction of Majoranas, due to the pseudo-magnetic field terms $(\propto\kappa)$. To introduce a hole at site $i$, we remove all the bonds to the site $i$, while keeping the spin itself.}
\label{fig1}
\end{center}
\end{figure}

{\it Model}: As a theoretical model of STM, we consider the setup shown in Fig.~\ref{fig1} (a): we place a target sample to realize Kitaev CSL on a metallic substrate, and allow conduction of the electric current from the STM tip to the substrate through the sample. We model this device by the Hamiltonian,
\begin{eqnarray}
\mathcal{H} = \mathcal{H}_{\rm sys} + \mathcal{H}_{\rm tip} + \mathcal{H}_{\rm sub} + \mathcal{H}^{\rm tip}_{\rm hyb} + \mathcal{H}^{\rm sub}_{\rm hyb}.
\end{eqnarray}
Here, $\mathcal{H}_{\rm sys}$ represents the Kitaev's system,
\begin{align}
\mathcal{H}_{\rm sys} = \mathcal{P}\mathcal{H}_{\rm K}\mathcal{P} + \mathcal{P}\mathcal{H}_{\rm hop}\mathcal{P}.
\label{eq:sys}
\end{align}
$\mathcal{H}_{\rm K}$ is the Kitaev's effective Hamiltonian with the pseudo-magnetic field term to realize the CSL phase,
\begin{align}
\mathcal{H}_{\rm K}=-J_{\rm K}\sum_{\langle i,j\rangle_{\alpha}}S_j^{\alpha}S^{\alpha}_{k} - 2\kappa\sum_{\langle j,k,l\rangle_{\alpha\beta}}S^{\alpha}_{j}S^{\beta}_{k}S^{\gamma}_{l},
\label{eq:KitaevHamiltonian}
\end{align}
where $S_j^{\alpha}$ is the spin-$1/2$ operator in fermionic representation, $S_j^{\alpha}\equiv\frac{1}{2}f^{\dag}_{j,s}\sigma^{\alpha}_{ss'}f_{j,s'}$, with $f^{\dag}_{j,s}$, the creation operator of a fermion at site $j$ and spin $s$. We focus on the case of hole injection into the half-filled Mott insulating state; $\mathcal{P}$ is the projection operator to exclude doublons.
$\mathcal{H}_{\rm hop}$ describes the motion of an injected hole.

$\mathcal{H}_{\rm tip}=\sum_{m\sigma}E_m^t\alpha^{\dag}_{m\sigma}\alpha_{m\sigma}$ and $\mathcal{H}_{\rm sub}=\sum_{l\sigma}E_l^s\beta^{\dag}_{l\sigma}\beta_{l\sigma}$ are the Hamiltonians of the tip and the substrate, modeled as simple non-interacting metals, characterized by the density of states, $\rho^{\rm t}(\varepsilon)$ and $\rho^{\rm s}(\varepsilon)$, respectively. $\mathcal{H}^{\rm tip}_{\rm hyb}=\sum_{m\sigma}v^{\rm t}_{m\sigma}(\alpha^{\dag}_{m\sigma}f_{i\sigma} + {\rm H.c.})$ describes the tunneling of electrons between the tip and Kitaev's system through the site $i$ in the Kitaev's system, and $\mathcal{H}^{\rm sub}_{\rm hyb}$ 
accounts for the tunneling process between the substrate and the system.

The STM current flowing from the tip to the sample can be written as~\cite{PhysRevLett.6.57,PhysRevB.31.805,PhysRevB.34.5947,PhysRevLett.103.206402}
\begin{eqnarray*}
I_i = -\frac{2\pi e}{\hbar}|v|^2\int\ d\omega\rho^{\rm t}(\omega + eV)\rho_i(\omega)[f(\omega + eV) - f(\omega)],
\label{eq:STMcurrent}
\end{eqnarray*}

where $\rho_i(\omega)$ is the local density of states of Kitaev's system at site $i$, in contact with the tip. For its derivation, see Supplemental material~\footnote{See Supplemental Material}. The tunneling amplitude is set to be constant: $v^{\rm t}_{m\sigma} = v$, for simplicity. We assume the Kitaev's system is in equilibrium with the substrate, and their common chemical potential, $\mu$, is tuned just above the top of the lower Hubbard band of Kitaev's Mott insulating state. Meanwhile, we set a voltage bias, $V$, between the tip and the system [Fig.~\ref{fig1} (b)], which drives the electric current between the system and tip, through the difference of distribution functions, $f(\varepsilon)=\frac{1}{e^{\beta(\varepsilon - \mu)} + 1}$.

If a simple constant density of states is assumed for the tip: $\rho^{\rm t}(\varepsilon)=\bar{\rho}$, the electric current can be simplified as $I_i=-\frac{2\pi e}{\hbar}|v|^2$$\bar{\rho}\int_{\mu}^{\mu-eV}\rho_i(\varepsilon)d\varepsilon$ at zero temperature. Accordingly, the differential conductance, $\frac{dI}{dV}$, gives direct information on the local density of states, $\rho_i(\mu - eV)$.

To address $\rho_i(\varepsilon)$ theoretically, we introduce the Hole Green's function, $g_{i\sigma}(t)\equiv-i\langle\Omega|f^{\dag}_{i\sigma}(t)f_{i\sigma}(0)|\Omega\rangle$, from which we can obtain the local density of states as $\rho_{i}(\varepsilon)=-\frac{1}{\pi}\sum_{\sigma}{\rm Im}g_{i\sigma}(\varepsilon)$. Here, $|\Omega\rangle$ represents the ground state of $\mathcal{H}_{\rm sys}$ at half filling of fermion (1 fermion per site), which is nothing but the ground state of Kitaev's Hamiltonian, $\mathcal{H}_{\rm K}$. 

To obtain the Green's function, $g_{i\sigma}(t)$, we need to diagonalize the Hamiltonian, $\mathcal{H}_{\rm sys}$, Eq.~(\ref{eq:sys}).
Among the two terms in Eq.~(\ref{eq:sys}), $\mathcal{H}_{\rm hop}$ represents the motion of an injected hole in the system. 
Below, we ignore $\mathcal{H}_{\rm hop}$, by assuming the influence of hole motion on the spin state is small, compared with magnetic interaction.
Under this assumption, the injected hole can be regarded as a site vacancy, and we can describe the intermediate state, $f_{i\sigma}(0)|\Omega\rangle$, as a superposition of the eigenstates of Kitaev's Hamiltonian, $\mathcal{H}_{\rm K}$, in the presence of a vacancy at site $i$.

To diagonalize $\mathcal{H}_{\rm K}$, we adopt Kitaev's Majorana representation of spin-$1/2$ operator, $S_j^{\alpha}=\frac{i}{2}c_jb_j^{\alpha}$, and rewrite the Hamiltonian, $\mathcal{H}_{\rm K}$, as
\begin{eqnarray}
\mathcal{H}_{\rm K}=\frac{i}{4}\sum_{jj'}c_jA_{jj'}c_{j'}=\sum_{m: \varepsilon_m>0}\varepsilon_{m}\bigl(\gamma^{\dag}_m\gamma_m-\frac{1}{2}\bigr),
\label{eq:Majorana_Hamiltonian}
\end{eqnarray}
where we have introduced complex fermion operators, $\gamma_m$. 
The presence of site vacancy does not spoil the integrability: one just needs to modify the Hamiltonian matrix, $A$.
Here, the first and second terms of $\mathcal{H}_{\rm K}$ in Eq.~(\ref{eq:KitaevHamiltonian}) are transformed into nearest- and next-nearest-neighbor hoppings of c-Majorana, $c_j$, as schematically shown in Fig.~\ref{fig1} (c). The b-Majoranas, $b_j^{\alpha}$, combine into gauge fluxes defined on each hexagon and appears as a phase in $A$. The $\pi$-vortex of gauge fluxes, called vison, increases the system energy by affecting the Hamiltonian matrix of c-Majorana, $A$~\cite{Lahtinen:2008aa,Lahtinen_2011}. An isolated vison accompanies one MZM in the CSL phase ($\kappa\not=0$), in the same mechanism as chiral $p$-wave superconductors~\cite{PhysRevLett.86.268} and Moore-Read fractional quantum Hall state~\cite{MOORE1991362}.

As for the dynamical response, one of the authors recently developed a technique to obtain a dynamical correlation function on a real-time basis~\cite{PhysRevB.98.220404,udagawa2019spectroscopy}. We apply this technique to obtain the Hole Green's function in the following form:
\begin{widetext}
\begin{eqnarray}
g_{i\sigma}(t) = -\frac{i}{2}\frac{\sqrt{{\rm det}(1 + e^{-(\beta-it)\cdot iA}e^{-it\cdot iA^i})} + (-1)^F\sqrt{{\rm det}(1 - e^{-(\beta-it)\cdot iA}e^{-it\cdot iA^i})}}
{\sqrt{{\rm det}(1 + e^{-\beta\cdot iA})} + (-1)^F\sqrt{{\rm det}(1 - e^{-\beta\cdot iA})}},
\label{eq:ChargeCorrelation}
\end{eqnarray}
\end{widetext}
where $(-1)^F$ is the physical fermion parity~\cite{pedrocchi2011physical,PhysRevB.98.220404}. $A^i$ and $A$ are the Hamiltonian matrices in Eq.~(\ref{eq:Majorana_Hamiltonian}) with and without a static hole at site $i$.
Specifically, $A^i$ is obtained by inputting zeros to the $i$-th row/column of matrix, $A$; As is the case with the previous studies on site vacancy~\cite{willans2010disorder,PhysRevB.84.115146,PhysRevB.90.035145,PhysRevB.94.235105,PhysRevB.98.220404,sanyal2020emergent}, we model the site vacancy at site $i$, by removing all the bonds connected to $i$. For the derivation of Eq.~(\ref{eq:ChargeCorrelation}), see Supplemental Material~\footnote{See Supplemental Material}.

Hereafter, we consider the system of $N\times N$ unit cells in the periodic boundary condition, with $N=48$. 
We consider two cases: (i) the flux free case; no hexagonal plaquette supports a vison, and (ii) the isolated vison case; a vison is introduced at the target site with a pair placed with maximal separation. We set $J_{\rm K}=1$ as a unit of energy, and $\beta=50$ for the computation of $g_{i\sigma}(t)$ in Eq.~(\ref{eq:ChargeCorrelation}).

\begin{figure}[h]
\begin{center}
\includegraphics[width=0.9\linewidth]{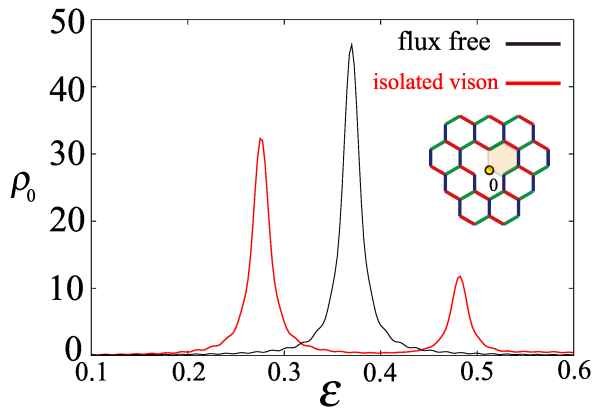}
\vspace{-10pt}
\caption{(Color online) Local density of state, $\rho_0(\varepsilon)$ for $\kappa=0.1$ at the neighboring site of a single vison (red), and in the flux free case (black).
The inset shows the target site $i=0$ with a yellow circle, and a vison with red shade.}
\label{fig2}
\end{center}
\end{figure}

{\it Results}:
Figure~\ref{fig2} represents our central result. Here, we set the tip at site $i=0$, and compare the local density of states, $\rho_0(\varepsilon)$ for the flux free case and the isolated vison case, for $\kappa=0.1$. 
In the latter case, we inject a hole on a site shared by the vison [Fig.~\ref{fig2} inset]. The two spectra show a clear difference. While the flux-free spectrum shows a single broad peak at $\varepsilon=\varepsilon_{\rm ff}\sim0.37$, the spectrum splits into two peaks in the presence of a vison. 
This clear difference can be accessed through the differential conductance, $\frac{dI}{dV}\propto\rho_0(\mu-eV)$, and thus it provides a clear diagnose of the presence of a vison.

To understand this spectral feature, we start with the flux free case. 
The physical origin of the single peak can be understood as follows.
Since we consider the limit where the charge motion is frozen, the spectrum is dominated by the magnetic excitation caused by the hole injection. 
The localized spin is removed by the hole injection, and this effectively disconnects three bonds.
Indeed, the resonant energy of flux-free spectrum, $\varepsilon_{\rm ff}$ is well estimated from the nearest-neighbor spin correlation, $3J_{\rm K}\langle s_i^{z}s_{i+z}^{z}\rangle\sim0.39$. 

\begin{figure}[h]
\begin{center}
\includegraphics[width=0.88\linewidth]{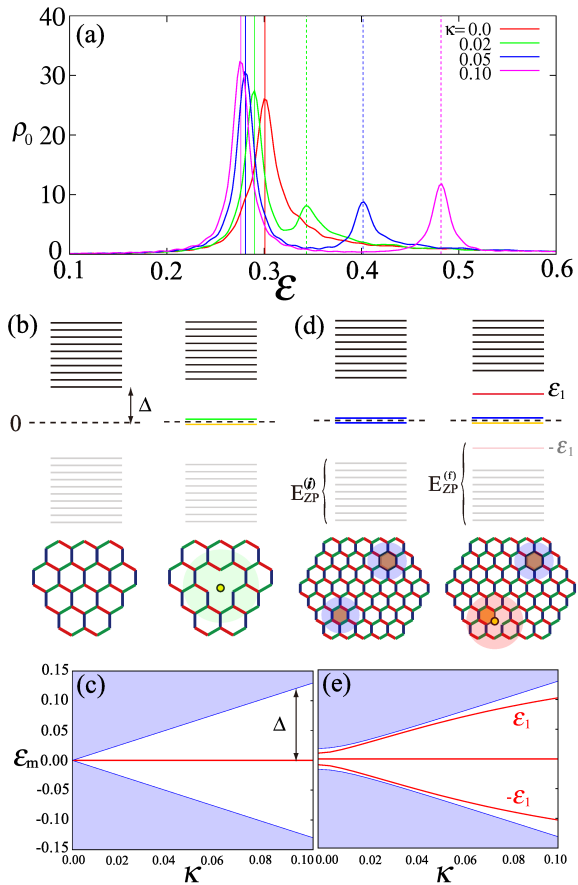}
\vspace{-10pt}
\caption{(Color online) (a)\ Local density of states at the neighboring site of single vison for $\kappa=0.0, 0.02, 0.05$ and $0.10$.  
The vertical lines are obtained from Eq.~(\ref{eq:resonant_energy}) with (solid line) $\delta\varepsilon^{(f)}=0$ and (dashed line) $2\varepsilon_1$, respectively. 
(b) and (d) show schematic energy levels (left) before, and (right) after the introduction of a hole, (b) in the flux free, and (d) in the isolated vison case. (b) Two zero modes appear after the hole injection; (yellow) one fictitious zero mode on the ``vacancy site" and (green) one unpaired Majorana, respectively. (d) Before the hole injection, two MZM exist as shown by blue lines, each attached to separate visons. After the hole injection, four low-energy modes appear; (yellow) a fictitious zero mode, (blue) a zero mode attached to a distant vison, and (red) the anti-bonding/bonding levels between the zero mode of target vison and the injected unpaired Majorana. 
(c) and (e) show numerically obtained energy levels after the hole injection (c) in the flux free, and (e) in the isolated vison case, corresponding to the right figures of (b) and (d).
The high-energy continuum is shown with blue shade. 
The low-energy modes are shown with red lines.
(c) Two modes are exactly degenerate at $\varepsilon=0$. (e) There exist nearly degenerate two modes at $\varepsilon\simeq0$, and the anti-bonding/bonding states at $\pm\varepsilon_1$.}
\label{fig3}
\end{center}
\end{figure}

Then, how does the presence of vison split this peak?
To see its origin, we plot $\rho_0(\varepsilon)$ in the presence of the vison as changing the pseudo-magnetic field term $\kappa$ in Fig.~\ref{fig3} (a).
For $\kappa=0$, the spectrum consists of only a single peak at $\varepsilon\sim0.3$, as in the flux free case.
However, for finite $\kappa$, it splits into one low-energy main peak and one high-energy subdominant peak, and the separation between the peaks becomes larger as increasing $\kappa$.

The low-energy main peak barely shifts from the original single peak position at $\kappa=0$, even when $\kappa$ is increased.
As explained above, the main peak can be explained by the release of the spin correlation energy.
Meanwhile, the subdominant peak quickly shifts to higher energy, when we increase $\kappa$. 

This subdominant peak reflects the information of the MZM carried by the vison. The key to understanding this is in the fact that the injected hole creates additional Majorana zero modes in the system, and interacts with the MZM if a vison is near the tip.
To clarify this, let us compare the energy level structures of the complex fermions ($\gamma_m$) in the absence and presence of the vison.
In Fig.~\ref{fig3} (b), we show the schematic energy level towers before and after introducing a hole in the flux free state.
A finite excitation gap $\Delta=\frac{3\sqrt{3}}{4}\kappa$, known as the Majorana gap, opens in the spectrum in the absence of the hole.
When the hole is introduced, two modes (red lines) appear inside the gap as depicted in Fig.~\ref{fig3} (c) for the whole range of $\kappa$.
They are the Majorana zero modes.
Among the two modes, one is physical and the other is fictitious.
The fictitious mode appears because we modeled the site vacancy by removing the bonds but keeping the site itself, and the fictitious mode sits on the hole site, where no spin actually exists.
In contrast, the other zero mode is real. It is an unpaired Majorana left in the bulk: Removal of one Majorana leaves an odd number of Majorana modes locally around the vacancy site, leading inevitably to a formation of one zero mode. It means the injection of a hole practically introduces a single unpaired Majorana in the bulk.

This unpaired Majorana leads to a nontrivial result in the presence of visons [Fig.~\ref{fig3} (d)]. 
To see this, let's assume that there were a pair of visons. 
If two visons are well separated, there exists a pair of Majorana zero modes, each attached to the visons, and the lift of their energies from zero is exponentially small determined by their separation.
Next, let us consider adding a hole near one of the vison. As we explained above, the hole adds one unpaired MZM, which then accompanies the MZM of the vison. The two MZM
form a bonding and anti-bonding pair whose energies are denoted as $\pm\varepsilon_1$ [Fig.~\ref{fig3} (e)].
The formation of this bonding/anti-bonding state opens a new channel of fermion excitations, which brings about the high-energy second peak in $\rho_0(\varepsilon)$.

To verify this scenario, let us look at the one-particle spectrum of Majorana Hamiltonian, Eq.~(\ref{eq:Majorana_Hamiltonian}).
In this representation, the resonant energy can be written as
\begin{eqnarray}
\omega^{i\to f}_{\rm res} = (E_{\rm ZP}^{(f)} - E_{\rm ZP}^{(i)}) + \delta\varepsilon^{(f)},
\label{eq:resonant_energy}
\end{eqnarray}
where the first term is concerned with the change of vacuum energy, $E_{\rm ZP}\equiv-\frac{1}{2}\sum_{m:\varepsilon_m>0}\varepsilon_m$, obtained from Eq.~(\ref{eq:Majorana_Hamiltonian}).
The introduction of a hole affects all the energy levels, resulting in the modification of $E_{\rm ZP}$. 
The second term, $\delta\varepsilon^{(f)}$, accounts for the changes in occupation numbers of fermions at each energy level.

On the basis of Eq.~(\ref{eq:resonant_energy}), we can attribute the low-energy main peaks to the vacuum-to-vacuum transition.
We obtain $\Delta E_0\equiv E_{\rm ZP}^{(f)} - E_{\rm ZP}^{(i)}$ for each $\kappa$, and plot them by vertical solid lines in Fig.~\ref{fig3} (a), which match the position of resonant peaks accurately.

In contrast, the high-energy second peaks involve fermion excitation. As mentioned above, at finite $\kappa$, the bonding between the injected unpaired MZM and the vison MZM results in discrete levels at $\pm\varepsilon_1$. Accordingly, we can again reproduce the position of second peaks in Fig.~\ref{fig3} (a) with the transition involving the fermionic excitation in the anti-bonding level, the vertical dashed lines, corresponding to $\Delta E_2\equiv (E_{\rm ZP}^{(f)} - E_{\rm ZP}^{(i)}) + 2\varepsilon_1$.

These two successful comparisons mean the separation between two resonant peaks scales with the bonding energy, $2\varepsilon_1$. As shown in Fig.~\ref{fig3} (e), $\varepsilon_1$ is of the order of Majorana gap, $\Delta$, proportional to $\kappa$. $\kappa$ is usually considered small, proportional to the cubic of a magnetic field, $h^3$\cite{kitaev2006anyons}, however, off-diagonal interactions may enhance its magnitude~\cite{PhysRevB.99.224409,takikawa2020topological}, and in fact, the estimate of $\Delta\sim10$K is proposed from the recent field-angle dependent specific heat~\cite{tanaka2020thermodynamic}. This large bonding energy of the CSL phase is in sharp contrast to the energy scale of vison-vacancy composites discussed in A and B phases, where the perturbative arguments result in $\sim10^{-6}J_z$ for A phase and $0.027J_{\rm K}$~\cite{willans2010disorder,PhysRevB.84.115146} for B phase, respectively. Compared with these cases, the stability of the vison-hole composite in the CSL phase is much more robust, promising clearer experimental identification of MZM.

\begin{figure}[h]
\begin{center}
\includegraphics[width=0.9\linewidth]{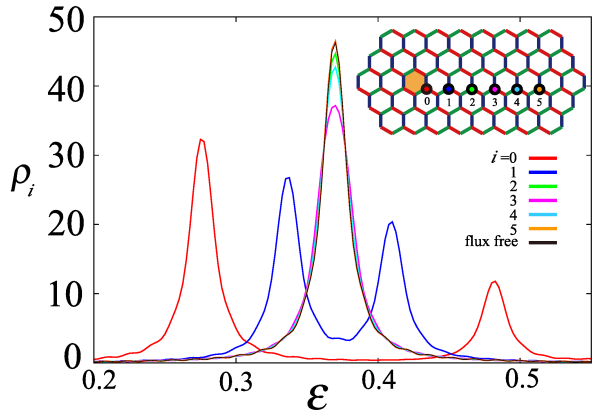}
\vspace{-10pt}
\caption{(Color online) Spatial dependence of local density of states, $\rho_i(\varepsilon)$. Site index convention is shown in the inset, together with the position of vison.
Away from the vison, $\rho_i(\varepsilon)$ approaches the form of flux-free case, represented by a black line.}
\label{fig4}
\end{center}
\end{figure}

What happens if the STM tip is near but away from the vison ? To clarify this, we inject the hole on sites $i$ away from the vison and plot the spectra $\rho_i(\varepsilon)$ in Fig.~\ref{fig4}.
As we move away from the hexagon supporting the vison, the two peaks quickly merge and the spectrum converges to the one in the flux-free case.
As a hole goes away from the vison, the coupling between Majorana zero modes becomes weaker, resulting in smaller bonding energy. The length scale of peak collapse depends on the excitation gap. This sensitive spatial dependence of $\rho_i(\varepsilon)$, relative to the location of vison, can be detected by sweeping the STM tip across the sample and provide useful information on the position of visons.

{\it Summary}: In summary, we have considered the hole injection into the Kitaev's chiral spin liquid phase, in the experimental setting of STM. We found that an injected hole effectively introduces a single Majorana into the system, and it forms a stable composite object with a precedent Majorana zero mode attached to a vison. The bonding energy of two Majoranas is of the order of Majorana gap of the CSL phase and is much larger than the binding energy of vison to a vacancy in A and B phases. The composite results in a double-peak structure of the local density of states, which can be observed through the differential conductance. The local density of states, with its characteristic magnetic field and spatial dependence, gives a good diagnose of the presence of Majorana zero mode attached to a vison. 

The use of STM has been recently proposed for the detection of edge states and fractional excitations~\cite{carrega2020tunneling,feldmeier2020local}.
In a broader scope, the local charge response of magnet may provide an access to ``nonlocal" information in terms of the magnetic degrees of freedom.
The recent rapid development of local charge sensitive probes may open an avenue to the measurement of nonlocal topological order, and to the long-awaited experimental identification of quantum spin liquids and their fascinating elementary excitations.

We deeply acknowledge Y. Matsuda, M. Knap, K. Damle, P. Wahl, P. A. McClarty and J. Knolle for helpful discussions. 
This work was supported by JSPS KAKENHI (Nos. JP15H05852 and JP16H04026), MEXT, Japan,
and JST CREST Grant No. JPMJCR19T3, Japan.

%

\widetext
\pagebreak

\renewcommand{\theequation}{S\arabic{equation}}
\renewcommand{\thefigure}{S\arabic{figure}}
\renewcommand{\thetable}{S\arabic{table}}
\setcounter{equation}{0}
\setcounter{figure}{0}
\setcounter{table}{0}

\begin{center}
\Large 
{Supplemental Material}
\end{center}
\section{Derivation of STM current}
We derive the STM current flowing from the tip to the sample, starting from the Hamiltonian of these two parts,
\begin{eqnarray}
\mathcal{H}=\mathcal{H}_{\rm tip} + \mathcal{H}_{\rm sys} + \mathcal{H}^{\rm tip}_{\rm hyb}.
\end{eqnarray}
$\mathcal{H}_{\rm sys}$ and $\mathcal{H}^{\rm tip}_{\rm hyb}$ are written as
\begin{eqnarray}
\mathcal{H}_{\rm sys}=\mathcal{P}\mathcal{H}_{\rm K}\mathcal{P},
\end{eqnarray}
\begin{eqnarray}
\mathcal{H}^{\rm tip}_{\rm hyb}=\sum_{m,\sigma}v^{\rm t}_{m\sigma}(\alpha^{\dag}_{m\sigma}f_{i\sigma} + {\rm H.c.}),
\end{eqnarray}
as introduced in the main text. 
We keep the system in equilibrium with the substrate with a common chemical potential, $\mu$. Meanwhile we control electrostatic potential of the tip, and set a voltage bias $V$ between the tip and the system.
The voltage bias affects the tip in two ways: it gives (i) a shift of one-particle energy, and (ii) a shift of chemical potential.
Concerning (i), the Hamiltonian of the tip is described by
\begin{eqnarray}
\mathcal{H}_{\rm tip}=\sum_m(E_m^t-eV)\alpha^{\dag}_m\alpha_m,
\end{eqnarray}
where the electron charge is set to be $-e$. Corresponding to (ii), we assume the chemical potential of the tip is given as, $\mu_{\rm tip}\equiv\mu - eV$.
Accordingly, the particle distribution is given by the shifted Fermi distribution function, $f(\varepsilon + eV) = \frac{1}{1 + e^{\beta(\varepsilon - \mu + eV)}}$. 

To obtain the current, we start with the equation of motion of electric charge at the target site, $i$,
\begin{eqnarray*}
\hat{I}_i\equiv -e\frac{d}{dt}\sum_{\sigma}f^{\dag}_{i\sigma}f_{i\sigma} = i\frac{e}{\hbar}\sum_{m,\sigma} v_{m\sigma}^t(\alpha^{\dag}_{m\sigma}f_{i\sigma} - f^{\dag}_{i\sigma}\alpha_{m\sigma}),
\end{eqnarray*}
which leads to the current expectation value, $I_i$, in terms of the non-equilibrium Green's function~\cite{kamenev2011field}.
\begin{eqnarray}
I_i = -\frac{e}{\hbar}\sum_{m\sigma}v_{m\sigma}^t\int\frac{d\varepsilon}{2\pi}{\rm Re}(g^{\rm K}_{st,im\sigma}(\varepsilon)).
\end{eqnarray}
Here, $g^{\rm K}_{st,im\sigma}(\varepsilon)$ is the Keldysh component of system-tip Green's function, defined from $g^{>}_{st,im\sigma}(t) = -i\langle f_{i\sigma}(t)\alpha^{\dag}_{m\sigma}\rangle, g^{<}_{st,im\sigma}(t) = i\langle \alpha^{\dag}_{m\sigma}f_{i\sigma}(t)\rangle$, and $g^{\rm K}_{st,im\sigma}(t) = g^{>}_{st,im\sigma}(t) + g^{<}_{st,im\sigma}(t)$.

we regard the tunneling Hamiltonian, $\mathcal{H}^{\rm tip}_{\rm hyb}$, to be small, and treat it with first-order perturbation theory.
From the standard double-path real-time perturbation theory~\cite{kamenev2011field}, we can attribute $g^{\rm K}_{st,im}(t)$ to the system and tip Green's functions as
\begin{eqnarray*}
g^{\rm K}_{st,im\sigma}(\varepsilon) = v_{m\sigma}^t(g^{\rm R}_{s,i\sigma}(\varepsilon)g^{\rm K}_{t,m\sigma}(\varepsilon) + g^{\rm K}_{s,i\sigma}(\varepsilon)g^{\rm A}_{t,m\sigma}(\varepsilon)).
\end{eqnarray*}
Here, $g^{\rm R}_{s,i\sigma}(\varepsilon)$ and $g^{\rm A}_{t,m\sigma}(\varepsilon)$ are retarded and advanced Green's function in the system and tip, respectively, and they are connected with spin- and level-resolved density of states, as
\begin{eqnarray*}
\rho_{i\sigma}(\varepsilon) = -\frac{1}{\pi}{\rm Im}g^{\rm R}_{s,i\sigma}(\varepsilon),\ \rho_{m\sigma}^t(\varepsilon + eV) = \frac{1}{\pi}{\rm Im}g^{\rm A}_{t,m\sigma}(\varepsilon).
\end{eqnarray*}
Meanwhile, in equilibrium, the Keldysh components are related to distribution function,
\begin{eqnarray}
g^{\rm K}_{s,i\sigma}(\varepsilon) = -2\pi i\rho_{i\sigma}(\varepsilon)(1-2f(\varepsilon))
\end{eqnarray}
\begin{eqnarray}
g^{\rm K}_{t,m\sigma}(\varepsilon) = -2\pi i\rho_{m\sigma}^t(\varepsilon+eV)(1-2f(\varepsilon + eV)).
\end{eqnarray}
Combining these equations together, and assuming simple tunneling matrix elements, $v^t_{m\sigma}\equiv v$, we obtain the electric current,
\begin{eqnarray*}
I_i = -\frac{2\pi e}{\hbar}|v|^2\int\ d\varepsilon\rho^{\rm t}(\varepsilon + eV)\rho_i(\varepsilon)[f(\varepsilon) - f(\varepsilon + eV)].
\end{eqnarray*}

\section{Derivation of hole Green's function}
As introduced in the main text, the hole Green's function of spin $\sigma=\pm1$ is written as
\begin{align}
g_{i\sigma}(t) &= -i\frac{{\rm Tr}[e^{-(\beta-it)\mathcal{H}_{\rm K}}f^{\dag}_{i\sigma}e^{-it\mathcal{H}_{\rm K}}f_{i\sigma}]}{{\rm Tr}\ e^{-\beta\mathcal{H}_{\rm K}}} = -i\frac{{\rm Tr}[e^{-(\beta-it)\frac{i}{4}c_kA_{kk'}c_{k'}}f_{i\sigma}e^{-it\frac{i}{4}c_kA_{kk'}c_{k'}}f^{\dag}_{i\sigma}]}{{\rm Tr}\ e^{-\beta\frac{i}{4}c_kA_{kk'}c_{k'}}}
\end{align}
Given the fermionic kinetic energy is frozen, the fermion annihilation operator, $f_{i\sigma}$, satisfies the following commutation relation with Hamiltonian:
\begin{align}
f^{\dag}_{i\sigma}e^{-it\frac{i}{4}c_kA_{kk'}c_{k'}}f_{i\sigma} &= e^{-it\frac{i}{4}c_kA^i_{kk'}c_{k'}}f^{\dag}_{i\sigma}f_{i\sigma} = e^{-it\frac{i}{4}c_kA^i_{kk'}c_{k'}}(\frac{1}{2} + \sigma S_i^z),
\end{align}
where $\frac{1}{2} + \sigma S_i^z$ is regarded as a projection operator on the state with spin $\sigma$ at site $i$. Accordingly, the Green's function can be transformed into
\begin{align}
g_{i\sigma}(t) = -i\frac{\sum_{\{W_p\}}{\rm Tr}_c[e^{-(\beta-it)\frac{i}{4}c_kA_{kk'}c_{k'}}e^{-it\frac{i}{4}c_kA^i_{kk'}c_{k'}}(i\sigma b_ic_i+\frac{1}{2})]}{\sum_{\{W_p\}}{\rm Tr}_c[e^{-\beta\frac{i}{4}c_kA_{kk'}c_{k'}}]}.
\label{eq:g}
\end{align}
Here, $\sum_{\{W_p\}}$ stands for the summation over the $Z_2$ flux configurations, $\{W_p\}$. ${\rm Tr}_c$ is the trace over $c$-fermions, implicitly involving the projection onto the physical fermion parity~\cite{pedrocchi2011physical}. In Eq.~(\ref{eq:g}), the term involving $b_j$ vanishes, as it changes the conserved flux sector. Accordingly, following the procedure in Ref.~\onlinecite{PhysRevB.98.220404}, we obtain
\begin{align}
g_{i\sigma}(t) &= -\frac{i}{2}\frac{\sum_{\{W_p\}}{\rm Tr}[e^{-(\beta-it)\frac{i}{4}c_kA_{kk'}c_{k'}}e^{-it\frac{i}{4}c_kA^i_{kk'}c_{k'}}]}
{\sum_{\{W_p\}}{\rm Tr}[e^{\beta\frac{i}{4}c_kA_{kk'}c_{k'}}]}\nonumber\\
&=-\frac{i}{2}\frac{\sum_{\{W_p\}}\sqrt{{\rm det}(1 + e^{-(\beta-it)\cdot iA}e^{-it\cdot iA^i})} + (-1)^F\sqrt{{\rm det}(1 - e^{-(\beta-it)\cdot iA}e^{-it\cdot iA^i})}}
{\sum_{\{W_p\}}\sqrt{{\rm det}(1 + e^{-\beta\cdot iA})} + (-1)^F\sqrt{{\rm det}(1 - e^{-\beta\cdot iA})}}.
\label{holegreenfunction}
\end{align}
Here, $(-1)^F$ is the physical fermion parity, which depends on the flux configuration. If we consider only a fixed sector of the flux configuration, we can omit the average over fluxes, and $g_{i\sigma}(t)$ can be simplified as Eq.~(5) in the main text.

\end{document}